\documentclass[aps,twocolumn,superscriptaddress,raggedbottom,showpacs,floatfix]{revtex4-1}
\usepackage[pdftex]{graphicx}
\usepackage{amssymb}
\usepackage{amsmath}
\usepackage{color}
\usepackage{hyperref}
\usepackage{textcomp}
\hypersetup{colorlinks=true,linkcolor=blue,urlcolor=blue,citecolor=blue}
\newcommand{\re}{\mathop{\rm Re}}
\newcommand{\im}{\mathop{\rm Im}}

\begin{document}

\title{Enhanced optical activity in hyperbolic metasurfaces}
\author{O.~V. Kotov}%
\email{oleg.v.kotov@yandex.ru}%
\affiliation{N.~L. Dukhov Research Institute of Automatics (VNIIA), 127055 Moscow, Russia}%
\affiliation{Institute of Microelectronics Technology and High Purity Materials RAS, 142432 Chernogolovka, Russia}%
\author{Yu.~E. Lozovik}% 
\email{lozovik@isan.troitsk.ru}%
\affiliation{N.~L. Dukhov Research Institute of Automatics (VNIIA), 127055 Moscow, Russia}%
\affiliation{Institute of Microelectronics Technology and High Purity Materials RAS, 142432 Chernogolovka, Russia}%
\affiliation{Institute for Spectroscopy RAS, 142190 Troitsk, Moscow, Russia}%
\affiliation{National Research University Higher School of Economics, 101000 Moscow, Russia}%

\begin{abstract}
	The optical activity of achiral nonmagnetic uniaxial metasurfaces caused by the extrinsic chirality arising from the mutual orientation of their anisotropy axis and the light plane of incidence is studied. The hyperbolic regime of the metasurfaces manifests in the amplification of the polarization rotation in the transmitted light and in the giant enhancement of the effect in the reflected light. The transition to this regime is frequency dependent and has the topological nature. The key role in the optical activity enhancement  belongs to the $\sigma$-near-pole and hyperbolic $\sigma$-near-zero regimes of the metasurfaces. The hyperbolic metasurfaces based on graphene strips or metal disks, and a black phosphorus thin film as a natural hyperbolic layer are considered. The efficiency of the predicted effects depends on the metasurfaces implementation and the role of a substrate. By varying the metasurface parameters and choosing the physical realizations it is possible to design such a hyperbolic metasurface polarizer for the THz, infrared, and visible working ranges. Our findings may help to improve metasurface polarizers by using them in the hyperbolic regime.
\end{abstract}

\maketitle

\section{Introduction} \label{Sec1}
Metasurfaces, the two-dimensional (2D) counterpart of metamaterials, have recently attracted a great attention due to their amazing properties, such as anomalous reflection and refraction, wavefront shaping and beam steering, subwavelength lensing, an impressive nonlinear response, and strong polarization conversion \cite{Rev_Holloway,Rev_Capasso,Zhao_Rev,ITMO_Rev,Chen_Rev}. In contrast to three-dimensional (3D) metamaterials, while possessing similar or superior functionalities, they allow one to remove the volumetric losses, to simplify the fabrication process, and to provide a full integration with planar optical devices. In the last few years, hyperbolic metasurfaces (HMSs), extremely anisotropic 2D structures behaving within the sheet as a dielectric along one direction and as a metal along the orthogonal one, following their 3D analog \cite{Smith_2003,Jacob_2012,Drachev_2013,RevHMM_ITMO,RevHMM_Jacob,HMM_twist,HMM_ITMO_gr,HMM_Zhang}, aroused a great interest owing to their strong confinement, negative refraction, and in-plane focusing of surface plasmon polaritons (SPPs), as well as an extremely large local density of states, Purcell factor, thermal emission, and photoluminescence polarization anisotropy \cite{Visible_HMS,Alu_PRL,Alu_OME,ITMO_PRB,ITMO_exp,Hanson_PRL,Liu_thermal,Alu_Rev,Alu2017,HMS_Lum}. The topology of HMS equal-frequency contours strongly depends on the frequency and changes from the elliptical to hyperbolic, thus giving rise to a tunable topological transition. The frequency of this transition is a function of HMS geometrical parameters and constituent material characteristics.

The phenomenon of optical activity can be caused not only by time-reversal symmetry breaking (external or internal magnetic fields), but also by mirror symmetry breaking (chirality). The 3D chiral nonmagnetic metamaterials \cite{Li_Rev,Oh_Rev,Wang_Rev} exhibit a differential absorption and transmission of left and right circularly polarized light (circular dichroism) and a rotation of the polarization state of light (circular birefringence) caused by the mirror asymmetry of constituent meta-molecules (intrinsic chirality). Besides, planar chiral structures with an in-plane intrinsic chirality show a 2D chiral response \cite{Zheludev_PRL2003} manifested in the asymmetric transmission of circularly polarized light with respect to the reversal of the direction of propagation (circular conversion dichroism) \cite{Zheludev_PRL2006,Zheludev_NL2008,Lederer_PRL2010,Alu_NC2012}. However, the optical activity induced by 3D intrinsic chirality is connected with the volumetric effect (the accumulation over the propagation distance) and thus can be large only for the transmitted light. What is more, the fabrication of intrinsically chiral metamaterials with 2D or 3D chiral responses is rather challenging. Nevertheless, one can achieve strong chiroptical effects using simply fabricated intrinsically achiral planar metamaterials (or metasurfaces) by tilting their symmetry axes with respect to the light plane of incidence. This phenomenon was first described by Bunn \cite{Bunn} and detected in nematic liquid crystals \cite{Williams_1968}, nonenantiomorphous crystals \cite{Futuma_1962,Chern_1970}, and layers with orientated molecules \cite{Appling_1986,Westphal_1989,Verbiest_1996} (for a detailed review see Refs.~\cite{Rev_Oloane,Barron_book,Rev_Konst}). Later, the effect when the chiroptical response in achiral structures is induced by the mirror asymmetry of the experimental arrangement was named pseudochirality  \cite{Sochava_book, Sersic} or \textit{extrinsic} chirality \cite{Plum,Plum2008,Plum_PRL,Plum_R}. For the extrinsic chirality, as well as for the intrinsic one, the circular dichroism is evidence of 3D chirality and circular conversion dichroism is a distinctive feature of 2D chirality. In the case of 3D extrinsic chirality, the mirror asymmetry arises from the oblique incidence on achiral but anisotropic metamaterials, which for different realizations has been studied experimentally \cite{Plum2008,Plum_PRL,Plum_R,APL_Ex3D,PRX_Ex3D,Sci_Ex3D2015,Sci_Ex3D} and theoretically \cite{OC_Th3D2012,OC_Th3D2013,Nsc_Th3D2016}, while 2D extrinsic chirality implies, first of all, a nonzero tilt angle between the metamaterial in-plane symmetry axes and the plane of incidence. For some specific in-plane configurations, 2D extrinsic chirality can be observed even at normal incidence \cite{Plum}, but usually for simple patterns (rectangles, holes, disks), to obtain circular conversion dichroism, both the tilt and incidence angles must be nonzero \cite{SciRep_Th2D2014,OE_Th2D2015,OME_Th2D,OE_Th2D2017,PhotRes_Th2D2017}. Recently, the optical activity (circular dichroism and birefringence) in achiral anisotropic 2D systems has been studied for graphene-based \cite{Gr_polar2015,Gr_polar2016,Gr_polar2017} or metallic \cite{Metal_polar2015,Metal_polar2017} metasurfaces, and for a natural 2D anisotropic layer, a black phosphorus (BP) thin film \cite{BP_polar}. Also, the asymmetry-induced chiroptical response of the thin hyperbolic metamaterial has been studied both theoretically and experimentally in Ref.~\cite{ENZ_polar} where the role of the $\varepsilon$-near-zero regime in the optical activity enhancement was revealed. Nevertheless, there was no in-plane asymmetry, so only the case of 3D chirality requiring an oblique incidence was presented. Finally, in Ref.~\cite{Gao_PRL} it was shown that by combining chirality and hyperbolicity it is possible to realize a photonic topological insulator with unidirectional surface electromagnetic waves topologically protected from backscattering. However, according to our knowledge, the hyperbolic chiroptical response for the 2D extrinsic chirality has not been studied yet. 

In this paper, we study the influence of the hyperbolic regime on the 2D extrinsic chiroptical response of achiral nonmagnetic uniaxial HMSs. In our setup we have a 2D chirality case when the polarization axis is tilted with respect to the HMS anisotropy axes (main axes). However, we consider only the polarization plane rotation which occurs due to the nonzero in-plane tilt angle even at normal incidence. To obtain circular conversion dichroism in our case one needs to break the relevant symmetries by using a substrate and oblique incidence \cite{ITMO_Sym}, which may become the subject of future work. Considering three different HMS realizations --- graphene strips,  metal disks, and a BP thin film as a natural HMS --- for all of them we demonstrate a giant enhancement of the polarization rotation which occurs exactly in the hyperbolic regime and is connected with the $\sigma$-near-zero resonances which are the 2D analog of $\varepsilon$-near-zero behavior in the 3D hyperbolic metamaterial studied in Ref.~\cite{ENZ_polar}. We also calculate the efficiency of the considered effects and the role of a substrate. 

\begin{figure}[b!]
	\centering
	\includegraphics[scale=0.37]{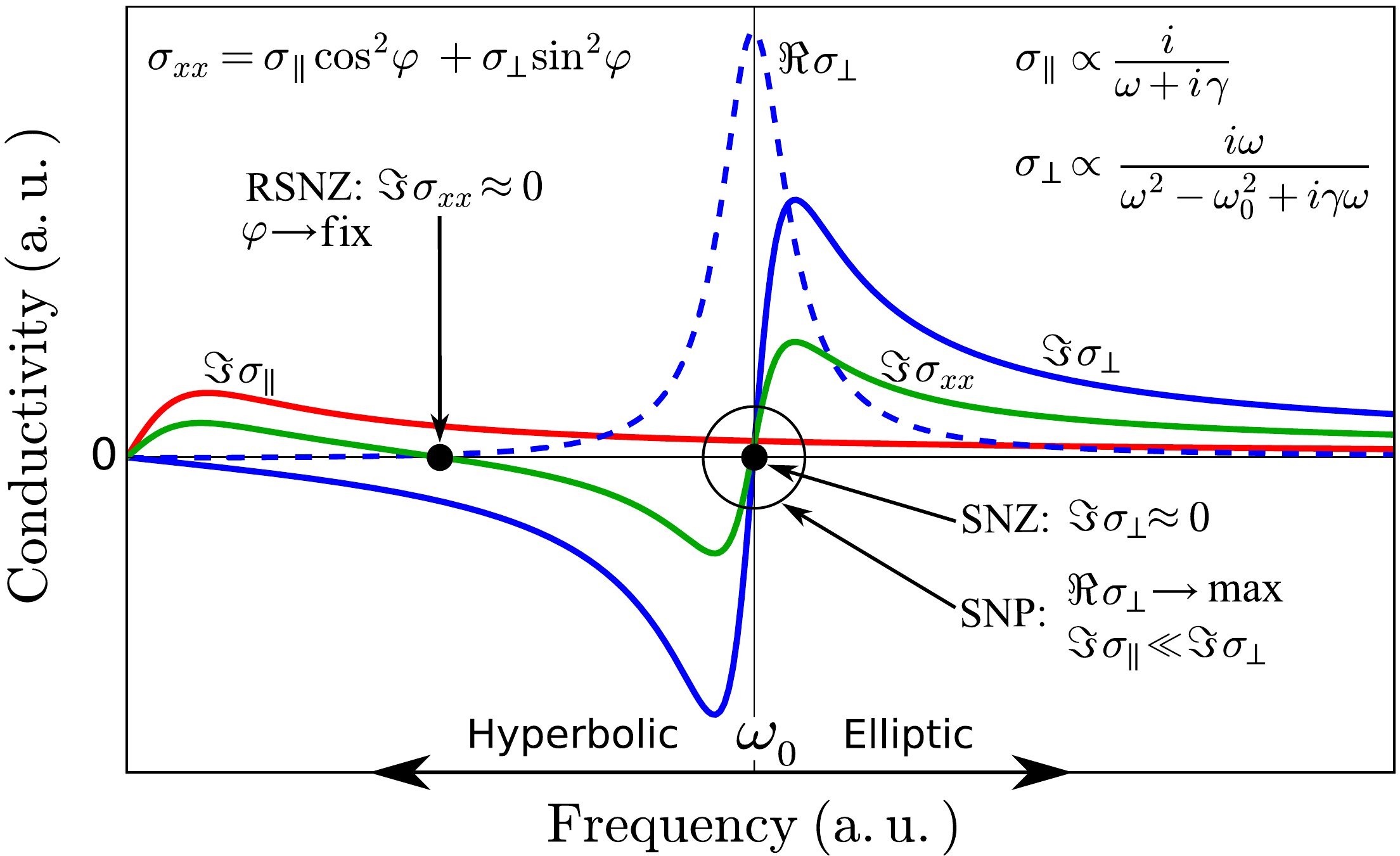} 
	\caption{\label{SNZ} Operating regimes of a model metasurface with Drude-like effective conductivity along the main axis ($\sigma_{\parallel}$) and Lorentz-like effective conductivity across it ($\sigma_{\perp}$). The topological transition from the elliptic to hyperbolic regimes occurs at the resonant frequency $\omega_0$. The black points denote the $\sigma$-near-zero (SNZ) regimes for $\sigma_{\perp}$ and $\sigma_{xx}$ [rotated-$\sigma$-near-zero (RSNZ) at fixed $\varphi$], and the black circle denotes the $\sigma$-near-pole (SNP) regime for $\sigma_{\perp}$.} 
\end{figure}

\section{Model} \label{Sec2}
\subsection{Effective conductivities and operating regimes} \label{Subs2a}
After the procedure of homogenization, which depends on the constitutive materials and geometry, one can describe a uniaxial metasurface by effective conductivities along ($\sigma_{\parallel}$) and across ($\sigma_{\perp}$) the main axis, and by an effective Hall conductivity ($\sigma^{\rm eff}_{\rm H}$) caused, e.g., by an external magnetic field. For a light with the plane of incidence is at an angle $\varphi$ (extrinsic chirality angle) to the main axis, a uniaxial metasurface can be characterized by the effective conductivity tensor rotated at the angle $\varphi$ \cite{ITMO_PRB,Alu_OME}:
\begin{widetext}
\vspace{-2em}	
\begin{align}
\begin{pmatrix}
\sigma_{xx} & \sigma_{xy} \\[0.5em]  \sigma_{yx} & \sigma_{yy}
\end{pmatrix} = 
\begin{pmatrix}
\sigma_{\parallel}\cos^2{\varphi} +\sigma_{\perp}\sin^2{\varphi} & \sigma^{\rm eff}_{\rm H} + \left(\sigma_{\parallel} -\sigma_{\perp}\right)\sin{2\varphi}/2 \\[0.5em] -\sigma^{\rm eff}_{\rm H} +\left(\sigma_{\parallel} -\sigma_{\perp}\right)\sin{2\varphi}/2 &  \sigma_{\parallel}\sin^2{\varphi} +\sigma_{\perp}\cos^2{\varphi}
\end{pmatrix}. \label{sig}
\end{align}
\vspace{-1em}
\end{widetext}
These effective conductivities describe the resonant interaction between the individual scatterers (meta-atoms) in a metasurface in corresponding directions. Usually, it is enough to account only for the dipole response at which the conductivities can be written in a Lorentzian form. Let us briefly discuss the typical operating regimes of such a metasurface. For this we consider a model configuration with Drude-like effective conductivity along the main axis [$\sigma_{\parallel}\propto i/(\omega+i\gamma)$] and Lorentz-like effective conductivity across it [$\sigma_{\perp}\propto i\omega/(\omega^2-\omega_0^2+i\gamma\omega)$]. In Fig.~\ref{SNZ} we plot these conductivities together with $\sigma_{xx}$, the component of the rotated tensor (\ref{sig}). At high frequencies the effective conductivities for both directions have a positive imaginary part $\im \sigma_{\parallel}>0,\;\im \sigma_{\perp}>0$, realizing the elliptic topology of the equal-frequency contours of the light dispersion in the metasurface \cite{ITMO_PRB}. At the resonant frequency $\omega_0$ the Lorentz-like perpendicular conductivity becomes negative, so at lower frequencies they have different signs of the imaginary parts $\im \sigma_{\parallel}>0,\;\im \sigma_{\perp}<0$, thus giving rise to a topological transition from the elliptic to the hyperbolic regime \cite{ITMO_PRB,Alu_PRL}. This topological transition frequency $\omega_0$ is the pole of $\sigma_{\perp}$-Lorentzian, so the behavior of a metasurface near $\omega_0$ can be called the $\sigma$-near-pole regime (by analogy with the $\varepsilon$-near-pole regime in 3D metamaterials \cite{Jacob_ENP,Jacob_ENPENZ,RevHMM_Jacob}). In this regime, $\re\sigma_{\perp}$ reaches a maximum and $\im\sigma_{\perp}$ exhibits a sign-changing resonance. Moreover, inside the SNP regime, there are also two $\sigma$-near-zero (SNZ) regimes: exactly at $\omega_0$: $\im\sigma_{\perp}=0$ (SNZ for $\sigma_{\perp}$), but near $\omega_0$: $\im\sigma_{\parallel}\ll\im\sigma_{\perp}$ (SNZ for $\sigma_{\parallel}$). These SNZ regimes are also analogous to the $\varepsilon$-near-zero regime in 3D metamaterials \cite{Engeta_ENZ,Jacob_ENPENZ,RevHMM_Jacob,ENZ_polar}. Furthermore, here we will consider the configurations with the extrinsic chirality (see Sec.~\ref{Sec1}) which causes the additional SNZ scenario for the rotated effective conductivities given by tensor (\ref{sig}). We will call this scenario the rotated-$\sigma$-near-zero (RSNZ) regime in which $\im\sigma_{xx}\approx0$ or $\im\sigma_{yy}\approx0$, while $\im \sigma_{\parallel}\neq0,\,\im \sigma_{\perp}\neq0$. Obviously, it can be realized only for specific chirality angles:  $\varphi=\pm\arctan\left(\sqrt{-\im \sigma_{\parallel}\rm/Im\,\sigma_{\perp}}\right)$ for $\im\sigma_{xx}=0$ and $\varphi=\pm\arctan\left(\sqrt{-\im \sigma_{\perp}\rm/Im\,\sigma_{\parallel}}\right)$ for $\im\sigma_{yy}=0$, which in the case of SPP excitation are the directions of their canalization \cite{Alu_PRL}. Notice that the RSNZ regime occurs only when $\im \sigma_{\parallel}$ and $\im \sigma_{\perp}$ have opposite signs, i.e., in the hyperbolic regime. Throughout this paper we will use the above terminology for the description of the optical activity effect in different metasurface realizations. It also must be underlined that nonlocalities arising due to the inhomogeneity of a metasurface associated with a meta-atom structure, or caused by the intrinsic spatially dispersive response of the constituent materials may sufficiently deform the hyperbolic equal-frequency contours and change the frequency range of the hyperbolic regime \cite{Alu_Nonloc}. Nevertheless, to demonstrate the optical activity enhancement in the hyperbolic regime, we find it enough to account only for the local response in the considered systems.
 
\vspace{-1em}
\subsection{Optical activity} \label{Subs2b}

When the light plane of incidence is tilted with respect to the main axis of an achiral nonmagnetic metasurface, the rotation of the effective conductivity tensor (\ref{sig}) gives the additional nondiagonal term $\left(\sigma_{\parallel} -\sigma_{\perp}\right)\sin{2\varphi}/2$, which corresponds to the extrinsic chirality. We will shorten the rotation of the polarization plane in the transmitted light by transmitted rotation (TR), and in the reflected light, by reflected rotation (RR). In the case with an external magnetic field, TR is called the Faraday effect and RR is the Kerr effect. Generally, in multipass experiments there can be a difference between the Faraday rotation and TR caused by a chirality: in the first case, the effect multiplies, but in the second one, it can be compensated. We will characterize these effects by the rotation angle and ellipticity $\theta_T$ and $\delta_T$ for the TR, and $\theta_R$ and  $\delta_R$ for the RR, correspondingly:  
\begin{align}
\theta_{\rm (T,R)} = \frac{1}{2}\arg\frac{(t,r)_-}{(t,r)_+},\,\,\,\delta_{\rm (T,R)} = \frac{\left| (t,r)_+\right|  - \left| (t,r)_-\right| }{\left| (t,r)_+ \right| + \left| (t,r)_-\right| }, \label{FK}
\end{align}
where for the linear $p$-polarized incident light the transmission and reflection coefficients of the elliptically polarized light scattered by a metasurface are written as $(t,r)_\pm = (t,r)_{pp}\pm i(t,r)_{sp}$, and for $s$-polarized incident light $(t,r)_\pm = (t,r)_{ss}\pm i(t,r)_{ps}$. Here, $t_{ij}$ and $r_{ij}$ ($i,j$ are $p$ or $s$) are, respectively, the transmission and reflection coefficients of $p$- and $s$-polarized incident light, which can be obtained from Maxwell's equations using the following boundary conditions for electric and magnetic fields ($\mathbf{E}$ and $\mathbf{H}$): 
\vskip -2em
\begin{equation}
\begin{array}{rcl}
	\left[\mathbf{n},\mathbf{H}_2\right] - \left[\mathbf{n},\mathbf{H}_1\right] &= &\, \displaystyle \frac{4\pi}{c}\hat{\sigma}\mathbf{E},  
	\\[1em]
	\left[\mathbf{n},\mathbf{E}_2\right] - \left[\mathbf{n},\mathbf{E}_1\right] &= & \,0, 
\end{array} \label{b1}
\end{equation}	
where $\mathbf{n}$ is a unit normal vector. After some algebra we obtain (compare with Refs.~\cite{Lakhtakia,Nikitin,Alu_OME}):
\begin{align}
	r_{ss} = \dfrac{\left(k_1-k_2-\sigma_{yy}\right)\left(\epsilon_1/k_1+\epsilon_2/k_2+\sigma_{xx}\right) + \sigma_{xy}\sigma_{yx}}{\left(k_1+k_2+\sigma_{yy}\right)\left(\epsilon_1/k_1+\epsilon_2/k_2+\sigma_{xx}\right) - \sigma_{xy}\sigma_{yx}}  \nonumber
	\\
	r_{sp} = \dfrac{2\sqrt{\epsilon_1}\sigma_{xy}}{\left(k_1+k_2+\sigma_{yy}\right)\left(\epsilon_1/k_1+\epsilon_2/k_2+\sigma_{xx}\right) - \sigma_{xy}\sigma_{yx}}  \nonumber
	\\
	r_{ps} = \dfrac{2\sqrt{\epsilon_1}\sigma_{yx}}{\left(k_1+k_2+\sigma_{yy}\right)\left(\epsilon_1/k_1+\epsilon_2/k_2+\sigma_{xx}\right) - \sigma_{xy}\sigma_{yx}}  \nonumber 
	\\
	r_{pp} = \dfrac{\left(k_1+k_2+\sigma_{yy}\right)\left(\epsilon_1/k_1-\epsilon_2/k_2-\sigma_{xx}\right) + \sigma_{xy}\sigma_{yx}}{\left(k_1+k_2+\sigma_{yy}\right)\left(\epsilon_1/k_1+\epsilon_2/k_2+\sigma_{xx}\right) - \sigma_{xy}\sigma_{yx}} \label{r}
	\\
	t_{ss} = \dfrac{2k_1\left(\epsilon_1/k_1+\epsilon_2/k_2+\sigma_{xx}\right)}{\left(k_1+k_2+\sigma_{yy}\right)\left(\epsilon_1/k_1+\epsilon_2/k_2+\sigma_{xx}\right) - \sigma_{xy}\sigma_{yx}}  \nonumber
	\\
	t_{sp} = \dfrac{2\sqrt{\epsilon_1}\sigma_{xy}}{\left(k_1+k_2+\sigma_{yy}\right)\left(\epsilon_1/k_1+\epsilon_2/k_2+\sigma_{xx}\right) - \sigma_{xy}\sigma_{yx}}  \nonumber
	\\
	t_{ps} = \dfrac{2\sqrt{\epsilon_2}k_1/k_2\,\sigma_{yx}}{\left(k_1+k_2+\sigma_{yy}\right)\left(\epsilon_1/k_1+\epsilon_2/k_2+\sigma_{xx}\right) - \sigma_{xy}\sigma_{yx}}  \nonumber
	\\
	t_{pp} = \dfrac{2\sqrt{\epsilon_1\epsilon_2}/k_2\,\left(k_1+k_2+\sigma_{yy}\right)}{\left(k_1+k_2+\sigma_{yy}\right)\left(\epsilon_1/k_1+\epsilon_2/k_2+\sigma_{xx}\right) - \sigma_{xy}\sigma_{yx}} \label{t}
\end{align}	
Here, $k_{1,2} = \epsilon_{1,2}\cos(\theta)$ are the normalized to $\omega/c$ transverse to the  two-dimensional electron gas (2DEG) wave vectors in the upper (with permittivity $\epsilon_1$) and lower (with permittivity $\epsilon_2$) media, $\theta$ is the angle of incidence counted from the normal vector, and all the components of the conductivity tensor $\sigma_{ij}$ are normalized to $c/4\pi$. For the isotropic 2DEG in an external magnetic field at normal incidence the Faraday (transmitted rotation) angle and ellipticity can be approximately expressed through the normalized Hall conductivity: $\theta_{\rm F}\simeq-\re \sigma_{\rm H}/2$ \cite{Nikitin,Ferreira} and $\delta_{\rm F}\simeq-\im \sigma_{\rm H}/2$ \cite{Ferreira}. Using Eqs.~(\ref{FK}) and (\ref{t}) we obtain similar expressions for the more general case when $\sigma_{xy}$ contains an additional chiral term (see (\ref{sig})). At zero external magnetic field ($\sigma_H=0$) they are written as 
\begin{equation} 
\begin{array}{rcl}  
	\theta_{\rm T}\simeq-\re \sigma_{xy}/2=\left(\re \sigma_{\perp}-\re \sigma_{\parallel}\right)\sin{2\varphi}/4, \displaystyle
	\\[1em]
	\delta_{\rm T}\simeq-\im \sigma_{xy}/2=\left(\im \sigma_{\perp}-\im \sigma_{\parallel}\right)\sin{2\varphi}/4.  
\end{array} \label{appr_F}
\end{equation}	
Equations~(\ref{sig}) and ~(\ref{appr_F}) have several interesting consequences. Firstly, at $\varphi=45^\circ$ a uniaxial metasurface becomes effectively isotropic with the conductivity $\sigma_{xx}=\sigma_{yy}=\left( \sigma_{\parallel}+\sigma_{\perp}\right)/2$ but having a nonzero nondiagonal component $\sigma_{xy}=\left( \sigma_{\parallel}-\sigma_{\perp}\right)/2$, which gives the optical activity effect. We find it rather amazing that an effectively isotropic metasurface can rotate light without magnetic field effects. Secondly, in the SNP regime (see Sec.~\ref{Subs2a}) $\re\sigma_{\perp}$ reaches a maximum, giving rise to a TR angle $\theta_{\rm T}$. Notice that an analogous $\varepsilon$-near-pole regime in 3D metamaterials leads to enhanced thermal emission in the corresponding narrow frequency range \cite{Jacob_ENP}. Thirdly, in the elliptical regime of a metasurface, when $\im \sigma_{\parallel}>0$ and $\im \sigma_{\perp}>0$, the TR ellipticity is determined by the difference between the conductivities along and across the main axis: 
\begin{align} 
	\left|\delta_{\rm T}^{\rm elliptic}\right|\sim\left|\left| \im \sigma_{\parallel}\right| -\left| \im \sigma_{\perp}\right|\right|.
	\label{del_ell}
\end{align}
While in the hyperbolic regime, when $\im \sigma_{\parallel}>0$ and $\im \sigma_{\perp}<0$, it is determined by their sum:
\begin{align}  
	\left|\delta_{\rm T}^{\rm elliptic}\right|\sim\left|\left| \im \sigma_{\parallel}\right| +\left| \im \sigma_{\perp}\right|\right|, \label{del_hyp}
\end{align}
which causes a rise of the ellipticity. Finally, using Eqs.~(\ref{FK}) and (\ref{r}) we also obtained that for the $p$-polarized incident light the RR angle caused by the extrinsic chirality at zero external magnetic field can be approximated by
\vskip -2em
\begin{align}  
	 \theta_{\rm R}\simeq\arctan\left(\im\sigma_{xy}/\im\sigma_{xx}\right). \label{thK_hyp}
\end{align}	
Notice that for the $s$-polarized incident light one should just replace in Eq.~(\ref{thK_hyp}) $\im\sigma_{xx}$ with $\im\sigma_{yy}$. In the SNZ regimes (see Sec.~\ref{Subs2a}) at $\im \sigma_{\perp}\approx0$ Eq.~(\ref{thK_hyp}) gives $\theta_{\rm R}\simeq\varphi$ and at $\im \sigma_{\parallel}\approx0$ it gives $\theta_{\rm R}\simeq\varphi-\pi/2$ (see Supplemental Material \cite{Suppl}). Moreover, according to Eq.~(\ref{thK_hyp}) for the $p$-polarized incident light the resonance in the RR is governed by the zeros of $\im\sigma_{xx}=\im \sigma_{\parallel}\cos^2{\varphi} +\im \sigma_{\perp}\sin^2{\varphi}\approx0$, which corresponds to the RSNZ regime for $\im\sigma_{xx}$ (see Sec.~\ref{Subs2a}). For $s$-polarized light also the RSNZ resonance takes place but for $\im\sigma_{yy}$. As we mentioned above, the RSNZ scenario occurs only in the hyperbolic regime, so if one excites SPP in HMS (e.g., with near-field radiation from a dipole), the RSNZ case will correspond to the canalization of hyperbolic SPP \cite{Alu_PRL}.

Thus the TR resonances are defined by the nondiagonal component of the effective conductivity, particularly, in the SNP regime $\theta_{\rm T}$ reaches a maximum and in the hyperbolic regime $\delta_{\rm T}$ gets amplification. Besides, the RR resonances are connected with the RSNZ hyperbolic regime. This is the key idea of the paper. We have demonstrated these effects of optical activity enhancement in the hyperbolic regime by examples of different realizations of HMSs which will be considered further. 

\section{Hyperbolic metasurface realizations} \label{Sec3}    
\subsection{Graphene strips} \label{Subs1}

\begin{figure*}[t]
	\centering
	\includegraphics[scale=0.26]{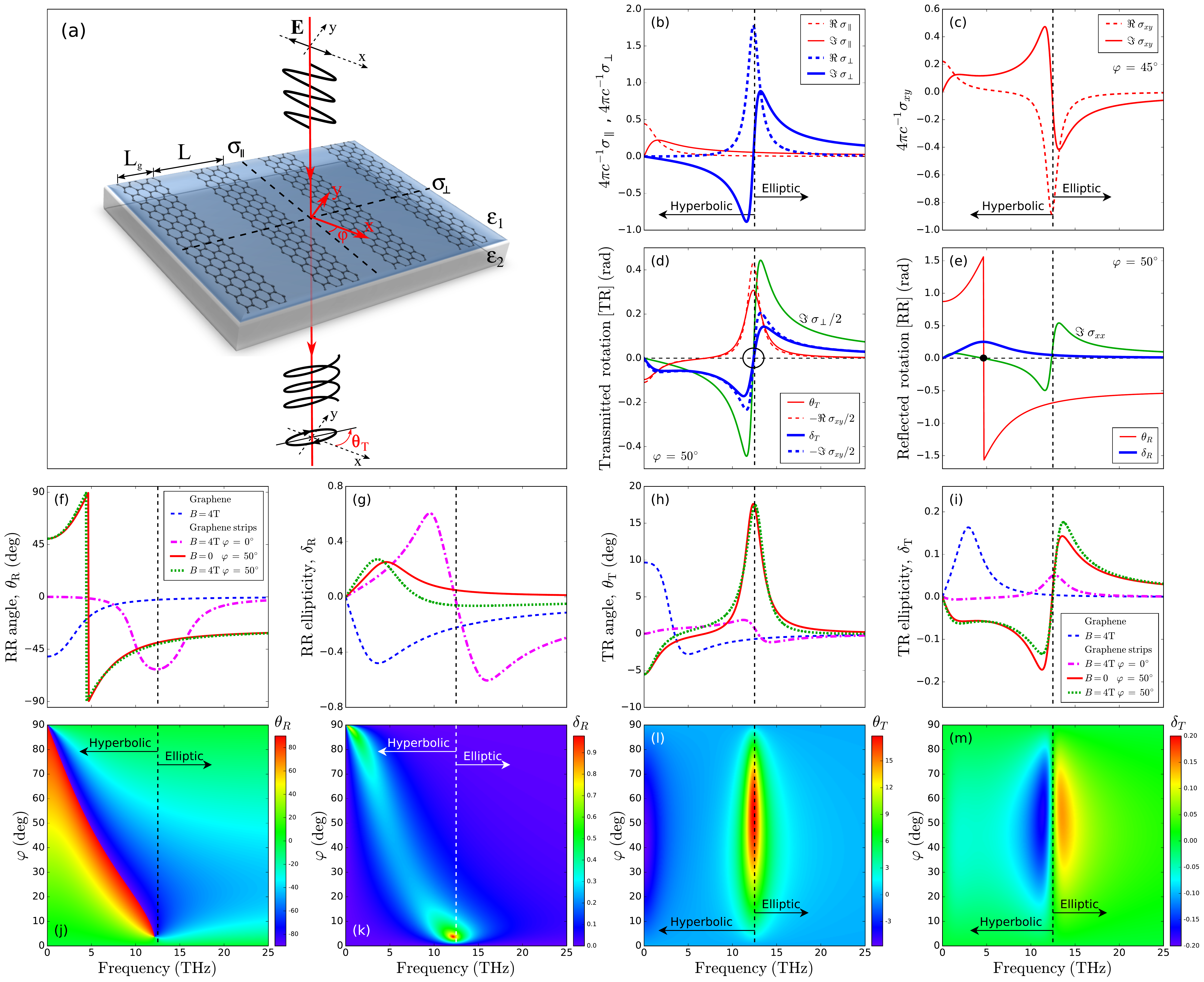} 
	\caption{\label{Gr} (a) Schematic representation of normally
		incident $p$-polarized light twisting by HMS based on graphene strips without magnetic field. The HMS located between two different media with the permittivities $\epsilon_1$ (above) and $\epsilon_2$ (below). The plane of incidence is at an angle $\varphi$ (extrinsic chirality angle) to the main axis, $\theta_{\rm T}$ is the angle of the polarization plane rotation of the transmitted  light. (b), (c) The effective conductivities: (b) along ($\sigma_{\parallel}$) and across ($\sigma_{\perp}$) the strips, (c) nondiagonal ($\sigma_{xy}$) at $\varphi=45^\circ$. (d) Transmitted rotation angle $\theta_{\rm T}$ and ellipticity $\delta_{\rm T}$ in comparison with $\sigma_{xy}$ at $\varphi=50^\circ$ showing the validity of the approximations given by Eq.~(\ref{appr_F}). The resonances in in the transmitted rotation are caused by the $\sigma$-near-pole regime (marked by black circle). (e) The same for the reflected rotation where the resonances are governed by the rotated-$\sigma$-near-zero regime (marked by black point) which occurs only in the hyperbolic regime [see Eq.~(\ref{thK_hyp})]. (f), (g) $\theta_{\rm R}$, $\delta_{\rm R}$ and (h), (i) $\theta_{\rm T}$, $\delta_{\rm T}$ for the various combinations of an external perpendicular magnetic field ($B$) and the extrinsic chirality at the fixed angle $\varphi$ (see the legends). (j), (k) $\theta_{\rm R}$, $\delta_{\rm R}$ and (l), (m) $\theta_{\rm T}$, $\delta_{\rm T}$ vs. the chirality angle and frequency. In (b)--(e) and (j)--(m) $B=0$. In (b)--(e) the conductivities are normalized to $c/4\pi$. In (b)--(m) the vertical dashed line indicates the topological transition frequency $\omega_0\approx12.4\,{\rm THz}$.  The parameters of the graphene strips HMS are set as follows: $L=1\,$\textmu m, $L_{\rm g}=0.5\,$\textmu m, $E_{\rm F}=0.2\,$eV, $v_{\rm F}=9.5\!\cdot\! 10^5\,$m/s, $\tau=0.1\,$ps, $\epsilon_1=\epsilon_2=1$.} 
\end{figure*}  
As the first example of HMS exhibiting an optical activity enhancement in the hyperbolic regime we have considered an array of graphene strips with the periodicity $L$ and the strip widths $L_{\rm g}$ on a dielectric substrate (see Fig.~\ref{Gr}). The homogenization of such a metasurface in the subwavelength approximation ($L\ll\lambda$) can be done using the effective medium theory based on the electrostatic approach \cite{Tretyakov_book}. According to this theory the graphene strips HMS along the main axes (along the strips) is similar to the continuous graphene with the conductivity $\sigma_{\rm g}$ and has an effective longitudinal conductivity $\sigma_{\parallel}=f_{\rm g}\sigma_{\rm g}$ with $f_{\rm g}=L_{\rm g}/L$ being the graphene filling factor. While across the main axes it behaves as an $LC$ circuit with the resonance defined by the effective transverse conductivity $\sigma_{\perp}=\sigma_{\rm g}\sigma_{\rm c}/\left(f_{\rm g}\sigma_{\rm c}+f_{\rm c}\sigma_{\rm g}\right)$, where $f_{\rm c}=1-f_{\rm g}$ is the filling factor of the free space between adjacent strips and $\sigma_{\rm c}$ is an equivalent conductivity associated with the near-field coupling between them. The grid impedance method combined with the approximate Babinet principle gives the following expression for the equivalent conductivity \cite{Luukkonen}:
\begin{align}  
\sigma_{\rm c}\approx-\frac{i\omega L}{\pi^2}\left(\frac{\epsilon_1+\epsilon_2}{2}\right)\ln\left[\csc\left(\frac{\pi f_{\rm c}}{2}\right)\right], \label{sig_c}
\end{align}
where $\epsilon_1$ and $\epsilon_2$ are the permittivities of the media above and below the strips. However, at nonzero Hall conductivity of graphene $\sigma_{\rm H}$ the components of the effective conductivity tensor (\ref{sig}) for the graphene strips HMS have a more complicated form \cite{Alu_Rev}: 
\begin{equation}
\begin{array}{rcl}
	\sigma_{\perp}=\sigma_{\rm g}\sigma_{\rm c}/\left(f_{\rm g}\sigma_{\rm c}+f_{\rm c}\sigma_{\rm g}\right),\quad \sigma^{\rm eff}_{\rm H}= f_{\rm g}\sigma_{\perp}\sigma_{\rm H}/\sigma_{\rm g}, \displaystyle
	\\[1em]
	\sigma_{\parallel}=f_{\rm g}\sigma_{\rm g}+f_{\rm g}\sigma_{\rm H}^2/\sigma_{\rm g}-\left(\sigma^{\rm eff}_{\rm H}\right) ^2/\sigma_{\perp}. \qquad
\end{array}  \label{sig_eff}
\end{equation}   
The hyperbolic regime comes from the sign-changing resonance in $\sigma_{\perp}$ which can be tuned by the varying HMS size parameters and graphene Fermi level, which allows one to work both in the THz and infrared (IR) ranges. In the THz range for a typical Fermi level $E_{\rm F}=0.2\,$eV one can neglect the impact of the interband transitions and describe graphene with the semiclassical conductivity tensor incorporating only intraband transitions in the local response approximation:
\begin{equation} 
\begin{array}{rcl} \displaystyle
	\sigma_{\rm g}=\frac{e^2E_{\rm F}}{\hbar^2\pi}\frac{i\left(\omega+i/\tau\right)}{\left(\omega+i/\tau\right)^2-\omega_c^2}, 
	\\[1em] \displaystyle \sigma_{\rm H}=\frac{e^2E_{\rm F}}{\hbar^2\pi}\frac{\omega_{\rm c}}{\left(\omega+i/\tau\right)^2-\omega_{\rm c}^2},
\end{array} \label{sig_g}
\end{equation} 
where $\tau$ is a relaxation time determined by the carrier mobility and $\omega_{\rm c}=eBv_{\rm F}^2/E_{\rm F}$ is the cyclotron frequency with $v_F$ being the Fermi velocity, and $B$ being a static magnetic field applied perpendicularly to the graphene sheet. Taking $L=1\,$\textmu m, $L_{\rm g}=0.5\,$\textmu m, $E_{\rm F}=0.2\,$eV, $v_{\rm F}=9.5\!\cdot\! 10^5\,$m/s, $\tau=0.1\,$ps, $\epsilon_1=\epsilon_2=1$, and using Eqs.~(\ref{sig}), (\ref{FK}),~(\ref{r}),~(\ref{t}), and (\ref{sig_c}) -- (\ref{sig_g}) for the normally
incident $p$-polarized light we plot in Fig.~\ref{Gr} the effective conductivities and optical activity characteristics for the graphene strips HMS. For these parameters we obtain a resonance in $\sigma_{\perp}$ at $\omega_0\approx12.4\,{\rm THz}$, which is the topological transition frequency of the system. At frequencies lower than it $\im \sigma_{\perp}$ becomes negative, giving rise to the hyperbolic regime [see Fig.~\ref{Gr}(b)]. Near $\omega_0$ there is a SNP regime (see Sec.~\ref{Subs2a}) which causes the TR resonances. It manifests in $\sigma_{xy}$  [Fig.~\ref{Gr}(c)] and consequently as a peak in $\theta_{\rm T}$ and a sign-changing resonance in $\delta_{\rm T}$, that can be seen in Fig.~\ref{Gr}(d), which shows the validity of the approximations given by Eq.~(\ref{appr_F}). As we expected [see Eqs.~(\ref{del_ell}),~(\ref{del_hyp})] in the hyperbolic regime due to the summation of the effective conductivities the ellipticity gets amplification [see Fig.~\ref{Gr}(m)]. For the RR at the angles $\varphi>5^\circ$  we obtained very strong sign-changing resonance in $\theta_{\rm R}$ (from $-90^\circ$ to $90^\circ$) and high peak in $\delta_{\rm R}$ (up to 1) which are always located in the hyperbolic regime range [see Figs.~\ref{Gr}(j),~\ref{Gr}(k)]. As we supposed [see Eq.~(\ref{thK_hyp})], these RR resonances are governed by the RSNZ hyperbolic regime [see Fig.~\ref{Gr}(e)]. In Figs.~\ref{Gr}(f),~\ref{Gr}(g) and Figs.~\ref{Gr}(h),~\ref{Gr}(i) we compared $\theta_{\rm R},\delta_{\rm R}$ and $\theta_{\rm T},\delta_{\rm T}$, respectively, for the following cases: \textit{(i)} continuous graphene in a perpendicular magnetic field, \textit{(ii)} graphene strips in a perpendicular magnetic field, \textit{(iii)} graphene strips without a magnetic field but with the extrinsic chirality at fixed $\varphi\neq0$, and \textit{(iv)} graphene strips with both a perpendicular magnetic field and extrinsic chirality. The first two cases for the Faraday effect were considered in Ref.~\cite{Nikitin}, where the results for the graphene strips in a perpendicular magnetic field were obtained with the rigorous mode-matching numerical approach. For the same parameters (the array "B" in Ref.~\cite{Nikitin}) we got very good agreement [see the dashed-dotted lines in Figs.~\ref{Gr}(h),~\ref{Gr}(i)], which shows the validity of the effective conductivity approximation in this frequency range. For the cases with extrinsic chirality we obtained much stronger resonances at the same frequency even without a magnetic field. Notice that in the case of graphene strips just in a magnetic field (without any type of chirality) both Faraday and Kerr resonances correspond to the excitation of magnetoplasmons which are localized on the strips, as for the Faraday effect obtained in Ref.~\cite{Nikitin}. However, the effective medium approach which we use in this work does not allow one to calculate the excitation of localized SPP, because it treats the strips array as a homogeneous layer. Nevertheless, the resonance in the perpendicular effective conductivity originates from the near-field coupling between adjacent strips [given by Eq.~\ref{sig_c}] which gives a pole in the $\sigma_{\perp}$-Lorentzian and corresponding SNP regime. Remarkably, the resonance frequency in $\sigma_{\perp}$ obtained with the effective medium approach coincides with the plasmon one obtained with the rigorous numerical approach in Ref.~\cite{Nikitin}.     

\subsection{Metal disks} \label{Subs2}
\begin{figure*}[t]
	\centering
	\includegraphics[scale=0.26]{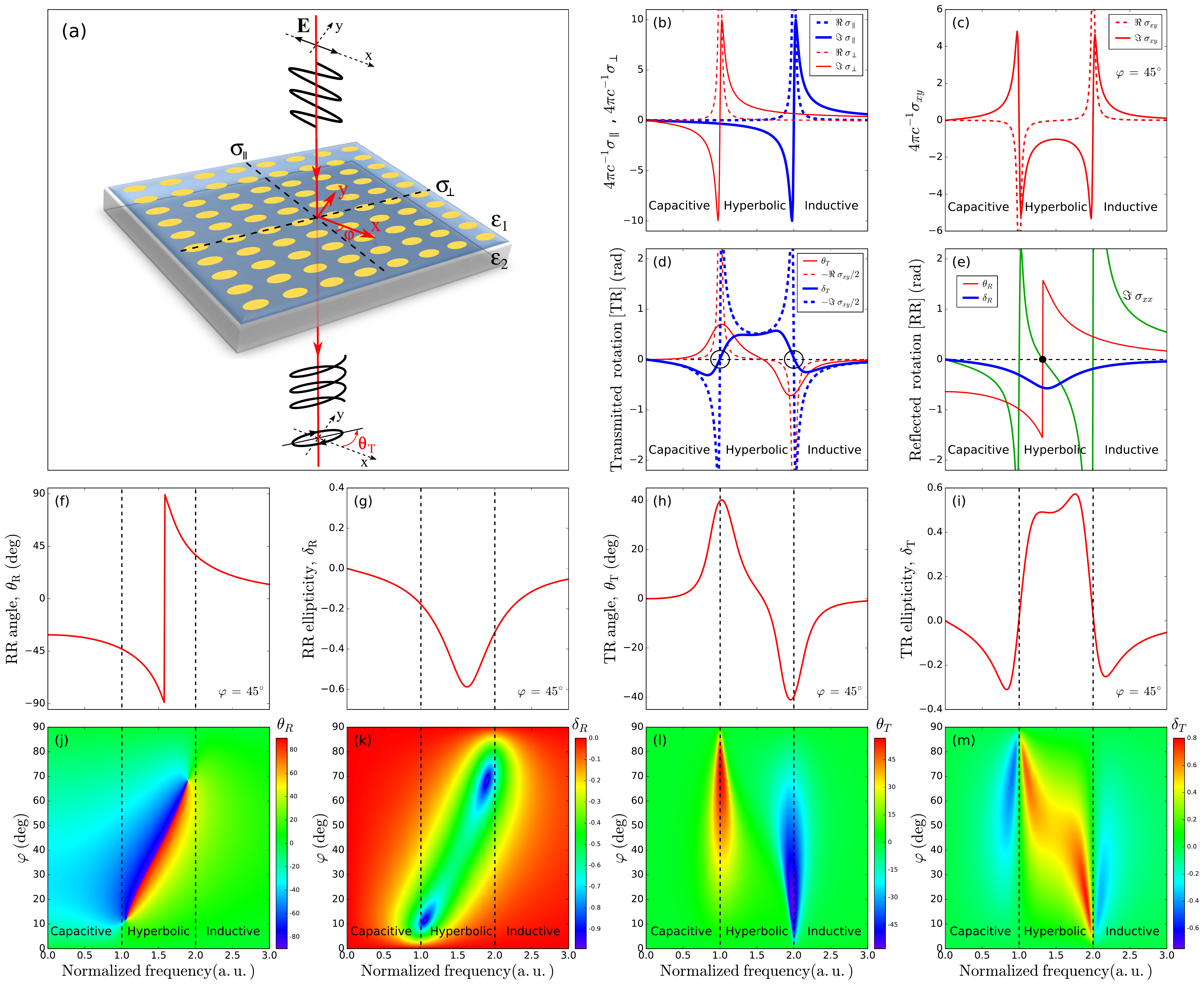} 
	\caption{\label{Met} The same as in Fig.~\ref{Gr} but for the metal disks HMS (2D lattice of anisotropic metal particles) with model effective conductivity [see Eq.~(\ref{sig_ITMO})] depending on the normalized frequency at $\tilde{\Omega}_{\perp}=1$, $\tilde{\Omega}_{\parallel}=2$, $\tilde{\gamma}=0.05$, $\epsilon_1=\epsilon_2=1$. The black circles in (d) show the $\sigma$-near-pole regime and the black point in (e) denotes the rotated-$\sigma$-near-zero regime.}
\end{figure*}
\begin{figure*}[t]
	\centering
	\includegraphics[scale=0.26]{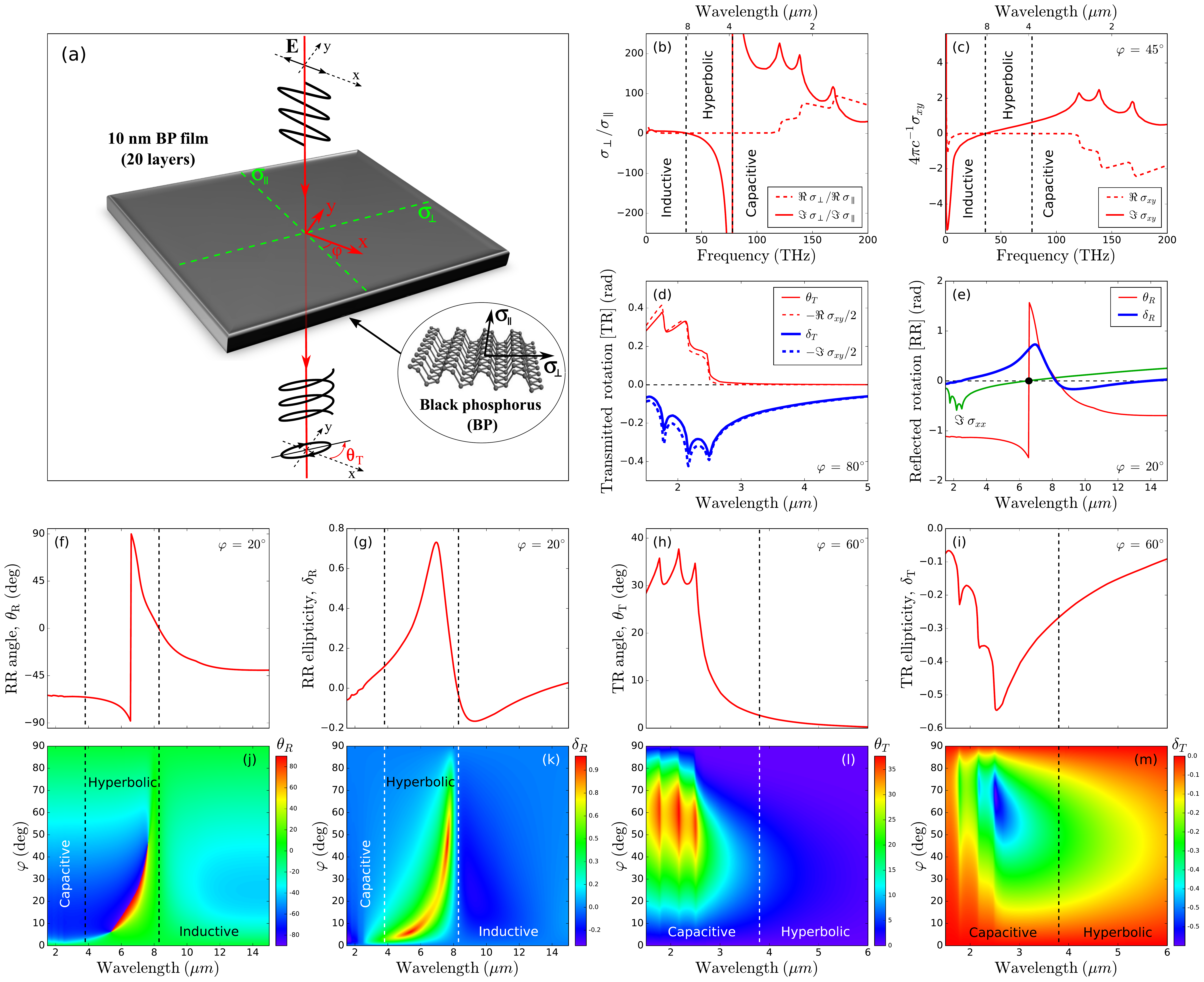} 
	\caption{\label{BP} The same as in Fig.~\ref{Gr} but for the natural hyperbolic layer, a black phosphorus 10-nm-thick film (20 atomic layers). $\sigma_{\parallel}$ (along the ridges) and $\sigma_{\perp}$ (across the ridges) are taken from Ref.~\cite{Alu_BP} at $E_{\rm F}=0.005\,$eV. The ratio $\im \sigma_{\perp}/\im \sigma_{\parallel}<0$ in (b) corresponds to the hyperbolic regime, and the black point in (e) denotes the rotated-$\sigma$-near-zero regime. 
	}
\end{figure*}
As the second example let us consider a 2D lattice of anisotropic metal particles for which the hyperbolic behavior recently has been predicted \cite{ITMO_PRB} and realized \cite{ITMO_exp}. The anisotropy can be caused by a different period in two perpendicular directions in the lattice or by the elliptical form of the particles (e.g., cylindrical gold disks with an elliptical base \cite{ITMO_exp}). Importantly, such HMSs can be engineered for the visible range \cite{ITMO_exp}. Following Ref.~\cite{ITMO_PRB} and accounting for the role of a substrate in the dipole and local response approximations, the effective conductivity tensor components can be written in a Lorentzian form:
\vskip -1.5em
\begin{align}
	\sigma_s=A_s\frac{ic}{4\pi}\frac{\omega}{\omega^2-\Omega_s^2/\epsilon_{\rm eff}+i\gamma_s\omega},\quad s\,=\,\parallel,\,\perp,  \label{sig_ITMO}
\end{align}
where $A_s$ is the normalizing factor with a dimension of rad/s and the effective permittivity of the  surrounding media $\epsilon_{\rm eff}=\left(\epsilon_1+\epsilon_2\right)/2$ as in Eq.~(\ref{sig_c}) accounts for the shift of the resonant frequencies $\Omega_s$ caused, e.g., by a substrate with $\epsilon_2$. Notice that for the 1D array of graphene strips Eqs.~(\ref{sig_c})--(\ref{sig_g}) give a similar Lorentzian resonance but only in the perpendicular direction ($\sigma_{\perp}$). Here for the 2D array of anisotropic metal particles we have two Lorentzians with different resonant frequencies in each of the two directions. This leads to the three different regimes of such a HMS: at low frequencies a capacitive one when both $\im \sigma_{\perp}$ and $\im \sigma_{\parallel}$ are negative, between the resonant frequencies a hyperbolic one when they have different signs, and at high frequencies an inductive regime when they are both positive as in a conventional anisotropic metal sheet. Assuming that the amplitudes and the bandwidth of the resonances are equal in both directions $A_{\parallel}=A_{\perp}=A$, $\gamma_{\parallel}=\gamma_{\perp}=\gamma$, and using Eqs.~(\ref{sig}), (\ref{FK}),~(\ref{r}),~(\ref{t}),~(\ref{sig_ITMO}) for the normally
incident $p$-polarized light at the normalized $\tilde{\omega}=\omega/A$, $\tilde{\Omega}=\Omega_s/A$, $\tilde{\gamma}=\gamma/A$, we plot in Fig.~\ref{Met} characteristics that are similar to the graphene strips HMS in Fig.~\ref{Gr}. Here we again confirm that TR resonances (in $\theta_{\rm T}$ and $\delta_{\rm T}$) are caused by the SNP regime [see Fig.~\ref{Met}(d)] and RR resonances (in $\theta_{\rm R}$ and $\delta_{\rm R}$) are governed by the RSNZ hyperbolic regime [see Fig.~\ref{Met}(e)]. Here RR resonances reach a maximum at the extrinsic chirality angles $10^\circ<\varphi<70^\circ$ [see Figs.~\ref{Met}(j),~\ref{Met}(k)]. Especially for such a HMS realization, where the hyperbolic area is between the resonant frequencies, one can see that the RR resonances occur exactly at the hyperbolic regime, but not just at low frequencies as can be thought for the graphene strips HMS [compare Figs.~\ref{Gr}(j),~\ref{Gr}(k) and Figs.~\ref{Met}(j),~\ref{Met}(k)]. The TR resonances at different normalized resonant frequencies $\tilde{\Omega}_{\parallel}$ and $\tilde{\Omega}_{\perp}$ have the opposite signs, moreover, one reaches its maximum at the chirality angles from low to middle $5^\circ<\varphi<65^\circ$ and the other, conversely, from high to middle $85^\circ>\varphi>35^\circ$ [see Fig.~\ref{Met}(l)]. Thus to obtain them both with a maximum level one should arrange the plane of incidence of a light at the middle angles to the main axes of the metal disks HMS, around $\varphi=45^\circ$. Notice that when the resonant frequencies are close enough, TR resonances (in $\theta_{\rm T}$ or $\delta_{\rm T}$) which occur exactly at the resonant frequencies may merge into a union peak lying in the hyperbolic area [see Fig.~\ref{Met}(m)]. The case of less difference between the resonant frequencies will be considered in Sec.~\ref{Sec4}.
	
\begin{figure*}[htp]
	\centering
	\includegraphics[scale=0.24]{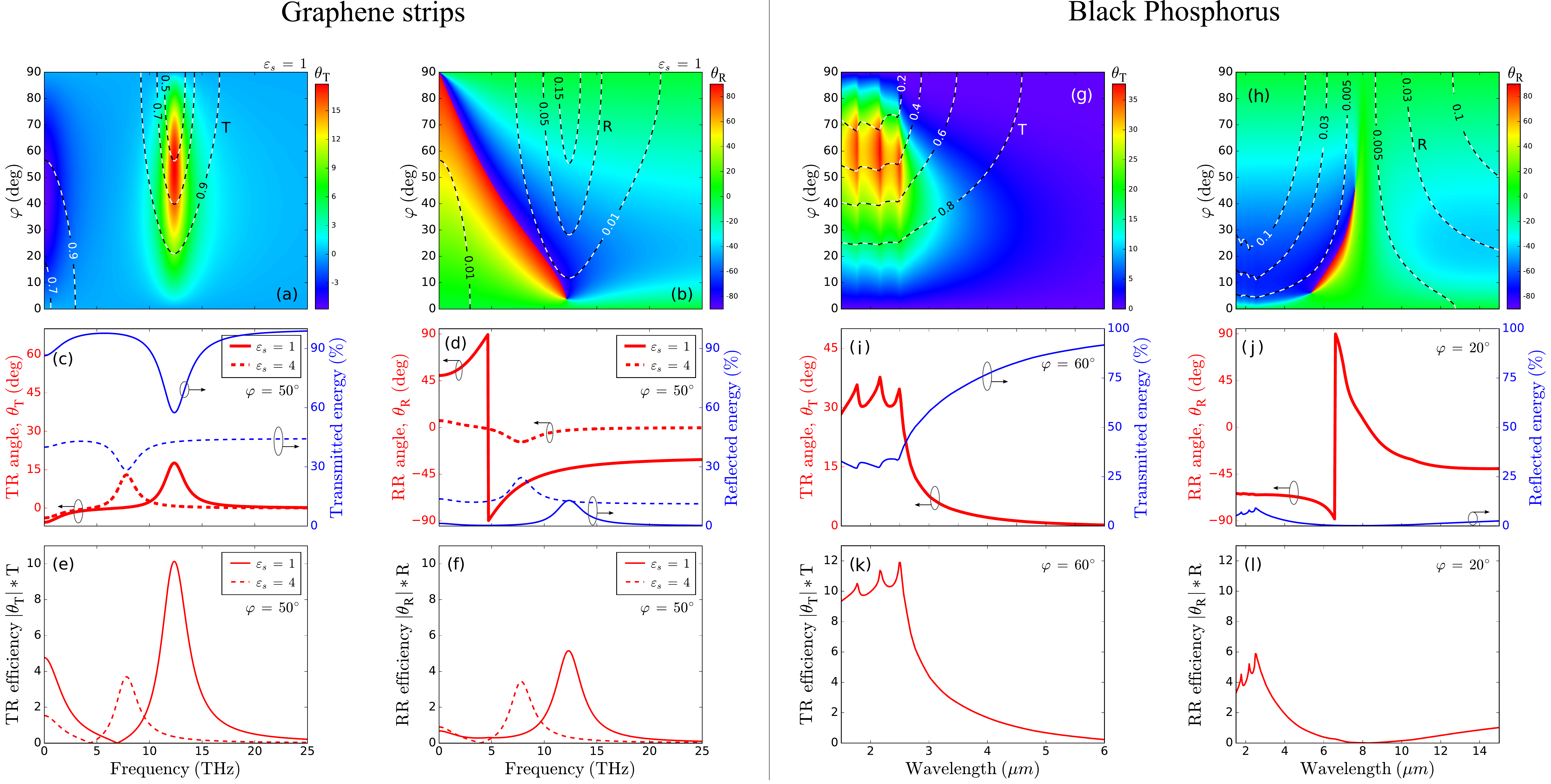} 
	\caption{\label{GrBP_eff} (a) and (b) The angles of the polarization plane rotation of the transmitted ($\theta_{\rm T}$) and reflected ($\theta_{\rm R}$) light, respectively, vs. the chirality angle $\varphi$ and frequency for the graphene strips HMS with parameters as in Fig.~\ref{Gr}. The levels show corresponding energy coefficients: transmittance ($\rm T$) and reflectance ($\rm R$) for the suspended system ($\epsilon_s=1$). (c) and (d) The same but at fixed $\varphi$ and also accounting for the substrate with $\epsilon_s=4$ (dashed lines). (e) and (f) The efficiencies at fixed $\varphi$ for the transmitted  $\left|\theta_{\rm T}\right|\cdot\rm T$ and reflected $\left|\theta_{\rm R}\right|\cdot\rm R$ rotations at $\epsilon_s=1$ (solid lines) and $\epsilon_s=4$ (dashed lines). (g)--(i) The same as in (a)--(f) but for the suspended black phosphorus thin film with parameters as in Fig.~\ref{BP}. 
	}
\end{figure*}

\subsection{Black phosphorus thin film} \label{Subs3}
In the previous examples we have considered artificial HMSs, but here let us study the nonmagnetic optical activity in a natural hyperbolic layer, a BP thin film. BP attracted a great deal of attention in the last years due to its strong in-plane anisotropy manifested both in linear \cite{Rev2015_BP,Rev2016_BP,Rev2017_BP,RevSmall2017_BP} and nonlinear \cite{Zhang1_BP,Zhang2_BP} optical responses. Unlike artificial metasurfaces, BP films possess a \textit{natural} in-plane anisotropy and strong tunability of the band gap and optical conductivity achieved by changing the number of layers, e.g., with laser pruning \cite{pruning_BP} (the BP band gap can vary from $\approx0.3\,eV$ in the bulk form to $\approx2.0\,eV$ in its monolayer form), or by an external electric field giving a giant Stark effect \cite{Stark_BP}, or with mechanical strain \cite{strain1_BP,strain2_BP}. All this provides additional control of the optical properties, but what is more, at a certain thickness the BP film becomes a natural hyperbolic layer. Particularly, as was predicted in Refs.~\cite{Alu_BP,Hanson_BP} a 10-nm-thick BP film (20 atomic layers) must exhibit the hyperbolic behavior at mid-IR frequencies. In terms of fabrication, on the one hand, artificial metasurfaces demand much more precision than natural anisotropic BP films, but on the other hand, BP thin films degrade rapidly in ambient conditions, so to stabilize them one needs to use passivation techniques covering BP with some protective layers \cite{protect_BP}. In any case, these layers are thin low-index dielectrics and have almost no influence on the optical activity effects, so for simplicity we will not consider them in our calculations. Notice that unlike in artificial metasurfaces, the nonlocal response in BP films has no inhomogeneity contribution associated with a meta-atom structure. However, the intrinsic spatially dispersive response in BP may sufficiently change the frequency range of the hyperbolic regime \cite{Alu_BP}. Nevertheless, for simplicity, as we have done for the previous HMS realizations, here we will account only for the local response in BP.

Taking from Ref.~\cite{Alu_BP} the conductivities $\sigma_{\parallel}$ (along the BP ridges) and $\sigma_{\perp}$ (across the ridges) calculated with Kubo formalism in the local response approximation at $E_{\rm F}=0.005\,$eV, and using Eqs.~(\ref{sig}), (\ref{FK}),~(\ref{r}),~(\ref{t}) for the normally
incident $p$-polarized light we plot in Fig.~\ref{BP} characteristics that are similar to the previous HMS realizations. In the BP film due to the strong anisotropy the conductivity along the ridges is much less than across them, so it is more convenient to plot the ratio $\im \sigma_{\perp}/\im \sigma_{\parallel}$ whose negative sign corresponds to the hyperbolic regime (Fig.~\ref{BP}(b)). Here also as in the metal disks HMS [see Sec.~\ref{Subs2}] there are three different regimes (inductive at $\omega<36\,\rm THz$, hyperbolic at $36\,\rm THz<\omega<78\,\rm THz$, and capacitive at $\omega>78\,\rm THz$), but their origin differs. While in artificial HMSs these regimes with different signs of the imaginary parts of the conductivities arise due to the resonances caused by the meta-atoms interaction, here they are caused by the different contributions of the interband electronic transitions in BP in two various directions (along and across the BP ridges). At low frequencies "inductive" (i.e., $\im \sigma<0$) intraband contributions dominate in all directions, and the system has a fully metallic behavior. At middle frequencies "capacitive" (i.e., $\im \sigma>0$) interband transitions begin to overcome the intraband impact first in the high-conductivity direction (across the ridges), and the system behaves as a dielectric across the ridges and as a metal along them (i.e., the system in the hyperbolic regime). Then at high frequencies the interband contributions start to dominate in all directions, and the system exhibits an entirely dielectric response. Also in contrast to artificial HMSs in such a BP film (without phonon or exciton resonances) there are no SNP regimes due to the absence of a resonant interaction between the meta-atoms inside the film. Nevertheless, for the BP film under consideration we get a strong enhancement of the TR angle and ellipticity in the capacitive regime [Figs.~\ref{BP}(l),~\ref{BP}(m)] which is caused by the large interband contributions in $\sigma_{\perp}$ at high frequencies [Figs.~\ref{BP}(b)--\ref{BP}(d)]. As for the HMSs considered above, here we also obtain strong RR resonances that are governed by the RSNZ hyperbolic regime [see Fig.~\ref{BP}(e)] at the extrinsic chirality angles $5^\circ<\varphi<60^\circ$ [Figs.~\ref{BP}(j),~\ref{BP}(k)].

\section{Efficiency and substrate effect} \label{Sec4} 
\begin{figure*}[t]
	\centering
	\includegraphics[scale=0.24]{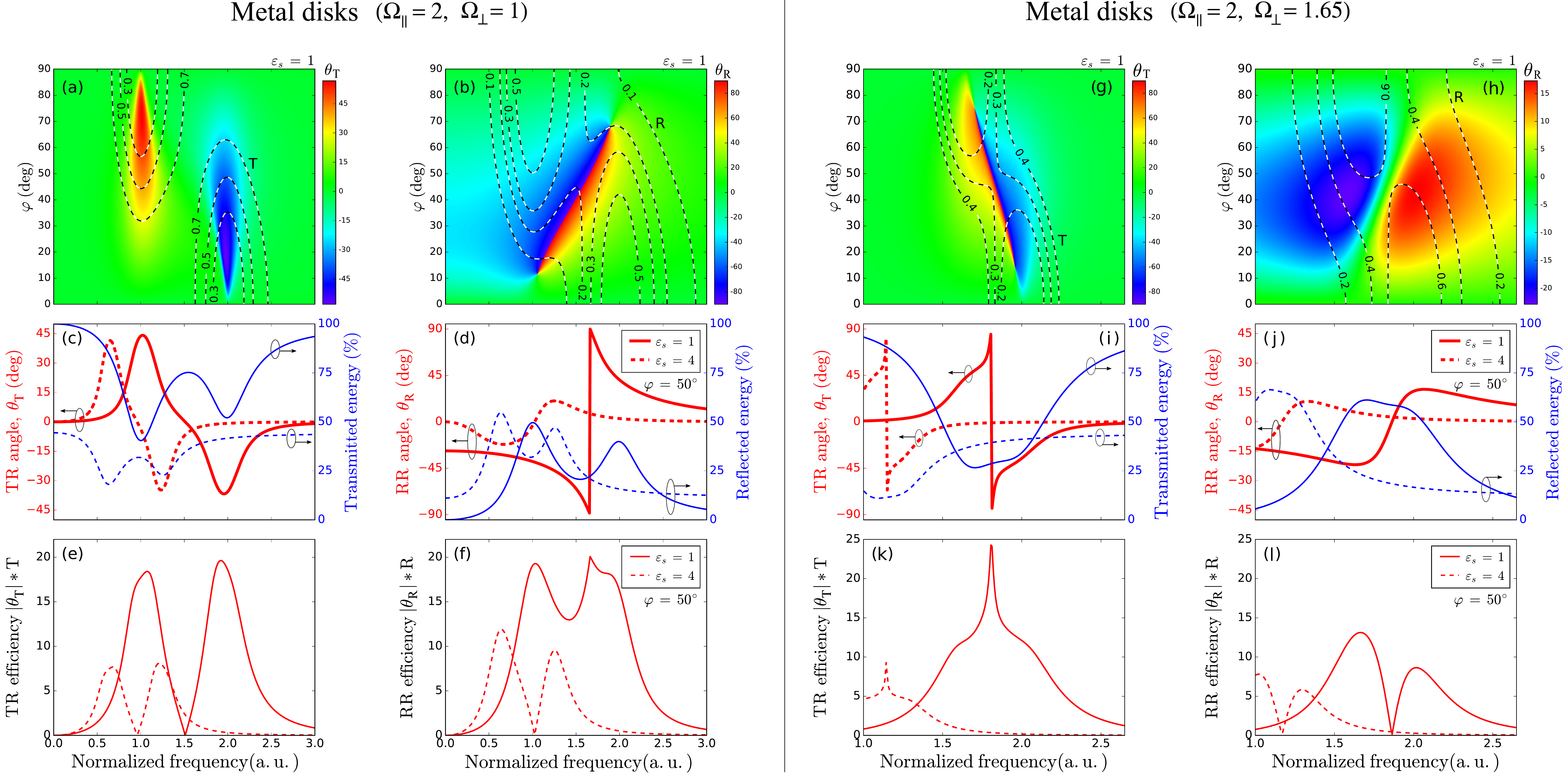} 
	\caption{\label{Met_eff} The same as in Fig.~\ref{GrBP_eff} but for the metal disks HMS described by the model effective conductivity [see Eq.~(\ref{sig_ITMO})] at normalized resonant frequencies $\tilde{\Omega}_{\perp}=1$, $\tilde{\Omega}_{\parallel}=2$ (a)--(f) and $\tilde{\Omega}_{\perp}=1.65$, $\tilde{\Omega}_{\parallel}=2$ (g)--(i); in both cases $\tilde{\gamma}=0.05$.  
	}
\end{figure*}
In the previous section we have focused on the magnitude of the rotation effects, but sometimes a strong rotation may be accompanied by a low percentage of the transmitted or reflected energy. Therefore, to estimate the efficiency of an optical rotator it is necessary to consider the product of the rotation angle and corresponding energy coefficient: $\left|\theta_{\rm T}\right|\cdot\rm T$ for the transmitted and $\left|\theta_{\rm R}\right|\cdot\rm R$ for the reflected light, where $\rm T=\left|t_{pp}\right|^2+\left|t_{sp}\right|^2$ and $\rm R=\left|r_{pp}\right|^2+\left|r_{sp}\right|^2$.

For the graphene strips HMS considered in Sec.~\ref{Subs1} the TR resonance with $\theta_{\rm T}\sim 15^\circ$ is accompanied by high transmittance $\rm T\sim50-80\%$ [see the levels in Fig.~\ref{GrBP_eff}(a)], but the RR resonance with $\theta_{\rm R}\sim 90^\circ$ corresponds to a very low reflectance $\rm R\sim1\%$ [see levels in Fig.~\ref{GrBP_eff}(b)]. As a result, the TR efficiency reaches a maximum ($\left|\theta_{\rm T}\right|\rm T\approx10\,[deg\cdot\%]$) at the resonant frequency [see solid lines in Figs.~\ref{GrBP_eff}(c), ~\ref{GrBP_eff}(e)], while the RR efficiency has the peak ($\left|\theta_{\rm R}\right|\rm R\approx5\,[deg\cdot\%]$) at frequency where the reflectance is the highest $\sim 15\%$ but the rotation is moderate $\sim 45^\circ$ [see solid lines in Figs.~\ref{GrBP_eff}(d),~\ref{GrBP_eff}(f)]. The thick substrate with $\epsilon_s=4$ [setting in Eqs.~(\ref{r}),~(\ref{t}),~(\ref{sig_c}) $\epsilon_1=1,\epsilon_2=4$] almost does not suppress the TR resonance and due to the change in the near-field coupling between adjacent strips [see Eqs.~(\ref{sig_c}),~(\ref{sig_eff})] only shifts it to a lower frequency. However, the substrate sufficiently decreases the transmittance, which leads to a reduction of the TR efficiency of more than two times [see dashed lines in Figs.~\ref{GrBP_eff}(c),~\ref{GrBP_eff}(e)]. Interestingly, it also results in the suppression of the RR efficiency: although the reflectance becomes higher, the rotation gets much more lower [see dashed lines in Figs.~\ref{GrBP_eff}(d),~\ref{GrBP_eff}(f)]. Notice that at $\epsilon_s=4$ we obtain that the RR resonance becomes the same as the TR one but with an opposite sign ($\theta_{\rm R}\approx-\theta_{\rm T}$), they occur at the same frequency at which also the reflectance approximately equals the transmittance ($\rm R\approx\rm T$), and all this results in  similar efficiency plots [compare the dashed lines in Figs.~\ref{GrBP_eff}(e) and \ref{GrBP_eff}(f)]. Thus the graphene strips HMS most efficiently rotates light in the transmitted mode at a resonant frequency lying in the THz range. Unfortunately, the giant RR resonance connected with the RSNZ hyperbolic behavior is accompanied by a very low reflectance and therefore does not contribute to the RR efficiency. Surprisingly, the substrate suppresses not only TR efficiency but also the RR one. 

For the BP thin film considered in Sec.~\ref{Subs3} we obtain [see Figs.~\ref{GrBP_eff}(g)-\ref{GrBP_eff}(l)] similar maximum values of the efficiencies ($\left|\theta_{\rm T}\right|\rm T\approx12\,[deg\cdot\%]$ and $\left|\theta_{\rm R}\right|\rm R\approx6\,[deg\cdot\%]$) which lie in the near-IR range (capacitive regime). Again here as for the graphene strips HMS, the giant RR resonance arising in the RSNZ hyperbolic regime is compensated by the negligible reflectance. 

The highest efficiency we obtain is for the metal disks HMS considered in Sec.~\ref{Subs2}. Moreover, here we study two cases: a moderate difference between normalized resonant frequencies (as in Fig.~\ref{Met} $\tilde{\Omega}_{\perp}=1$, $\tilde{\Omega}_{\parallel}=2$) and a small difference with $\tilde{\Omega}_{\perp}=1.65$, $\tilde{\Omega}_{\parallel}=2$. In the first case we get the TR resonances with $\theta_{\rm T}\sim 45^\circ$ accompanied by high transmittance $\rm T\sim50-70\%$ [see the levels in Fig.~\ref{Met_eff}(a)] and the RR resonance with $\theta_{\rm R}\sim 90^\circ$ corresponds to the reflectance $\rm R\sim20-30\%$ [see the levels in Fig.~\ref{Met_eff}(b)] which is much higher than in the graphene strips HMS. This results in two peaks of the TR and RR efficiencies ($\left|\theta_{\rm T}\right|\rm T\approx\left|\theta_{\rm R}\right|\rm R\approx20\,[deg\cdot\%]$ see Figs.~\ref{Met_eff}(e),~\ref{GrBP_eff}(f)) which are two (for TR) and four (for RR) times larger than the same characteristics for the graphene strips HMS or the BP thin film. As for the graphene strips HMS here a substrate shifts the resonant frequencies and suppresses the efficiencies. However, even with the substrate, here the RR efficiency ($\approx10[\rm deg\cdot\%]$) is two times greater than for the suspended graphene strips HMS or BP thin film. In the second case with a small difference between normalized resonant frequencies, the two TR resonances with opposite signs of the rotation angles merge into one sign-changing resonance [see Fig.~\ref{Met_eff}(g)]. This gives the amplification of the rotation at the union resonance to the highest values $\theta_{\rm T}\sim 90^\circ$ and the decrease of the transmittance to the moderate levels $\rm T\sim20-30\%$. As a result, the TR efficiency has a union enhanced maximum with $\left|\theta_{\rm T}\right|\rm T\approx25\,[deg\cdot\%]$ [see Fig.~\ref{Met_eff}(k)]. However, unlike the TR resonance, the RR one becomes significantly depressed, which results in the decreased RR efficiency $\left|\theta_{\rm T}\right|\rm T\approx13\,[deg\cdot\%]$ [see Fig.~\ref{Met_eff}(l)]. Thus using the metal disks HMS it is possible to obtain a highly effective strong rotation ($\theta\sim 90^\circ$) both for the reflected (in  the case $\tilde{\Omega}_{\perp}=1$, $\tilde{\Omega}_{\parallel}=2$) and transmitted (in the case $\tilde{\Omega}_{\perp}=1.65$, $\tilde{\Omega}_{\parallel}=2$) light. These strong resonances occur in the hyperbolic regime, which can be realized in the visible range \cite{ITMO_exp}. 

Notice that unlike magneto-optical Faraday and Kerr effects, where the rotation can be accumulated in the multipass experiment, here for the extrinsic chirality angles $0^\circ<\varphi<90^\circ$ the rotation has one sign, and for the angles $90^\circ<\varphi<180^\circ$ the opposite one, and thus it can be canceled. In the Supplemental Material \cite{Suppl} we provide  a full angle distribution of $\theta_{\rm T}$ and $\theta_{\rm R}$ for the graphene strips and metal disks HMSs which shows that in the multi-pass experiment the rotation periodically can compensate itself. After repeated interactions of HMS with light the angles $\theta_{\rm T}$ or $\theta_{\rm R}$ will be added to the angle $\varphi$ making it sometimes lie in the first or third quadrants and sometimes in the second or fourth quadrants where the sign of rotation is opposite. Thus it is better to observe the predicted effects in achiral nonmagnetic HMSs using a single-pass experimental arrangement.

We remark also that a strong nonlinear response of HMS constituent materials (e.g., in graphene \cite{Mikhailov_NL,Zhang_NL}) probably may lead to an even greater enhancement of optical activity. However, it would require strong laser illumination, while here we predict the effects which can be observed even in the linear response without using high-power light sources.   

\section{CONCLUSION}
In summary, using the effective conductivity approach we have studied the optical activity of achiral nonmagnetic uniaxial metasurfaces caused by the extrinsic chirality arising from the mutual orientation of their anisotropy axis and the light plane of incidence. We have obtained that a frequency dependent topological transition to the hyperbolic regime in such metasurfaces manifests in the amplification of the polarization rotation in the transmitted light (namely, the ellipticity rises) and in the giant enhancement in the reflected light (especially, the rotation angle grows). Moreover, while the enhancement for the transmitted light rotation is caused by the $\sigma$-near-pole resonances, for the reflected light it is governed by the rotated-$\sigma$-near-zero behavior which occurs only in the hyperbolic regime. By varying the metasurface parameters and choosing the physical realizations it is possible to design such a hyperbolic metasurface polarizer for the THz, infrared, and visible working ranges. The efficiency of the effect depends on the HMSs implementation and can be suppressed by a substrate. The graphene strips HMS and black phosphorus thin film provide a giant enhancement of the optical activity (reflected light rotation $\sim90^\circ$) in the THz and mid-IR ranges, respectively, but with low efficiency (corresponding reflectivity $\sim1\%$). We find the 2D lattice of anisotropic metal particles is the best HMS to observe enhanced optical activity (rotation angles $\sim90^\circ$ with corresponding energy coefficients $\sim25\%$) both for the reflected and transmitted light in the visible range. The predicted effects can be observed experimentally by using standard polarization analyzers. These results may help to improve metasurface polarizers by using them in the hyperbolic regime, thus making an important contribution to modern flat optics.

\vspace{1em}
\section*{Acknowledgments}

The authors are grateful to \mbox{M. A. Remnev} and \mbox{S. V. Remizov} for useful discussions. The work was supported by the Russian Science Foundation (Grant 17-12-01393). 

%\bibliographystyle{apsrev4-1}%
%\bibliography{bibHMS}%
%

\widetext
\pagebreak
%\vspace{1.5em}
\begin{center}
	\textbf{\large Supplemental Material: Enhanced optical activity in hyperbolic metasurfaces}
\end{center}

\begin{center}
	O.~V. Kotov${}^{1}$ and Yu.~E. Lozovik${}^{1,2,3}$
	
	\it{${}^1$N.~L. Dukhov Research Institute of Automatics (VNIIA), 127055 Moscow, Russia}
	
	\it{${}^2$Institute for Spectroscopy, Russian Academy of Sciences, 142190 Troitsk, Moscow, Russia}
	
	\it{${}^3$National Research University Higher School of Economics, 101000 Moscow, Russia}
\end{center}

\setcounter{figure}{0}
\setcounter{equation}{0}
\setcounter{section}{0}

\section{Sigma-near-zero and sigma-near-pole resonances} 
\begin{figure*}[b]
	\centering
	\includegraphics[scale=0.36]{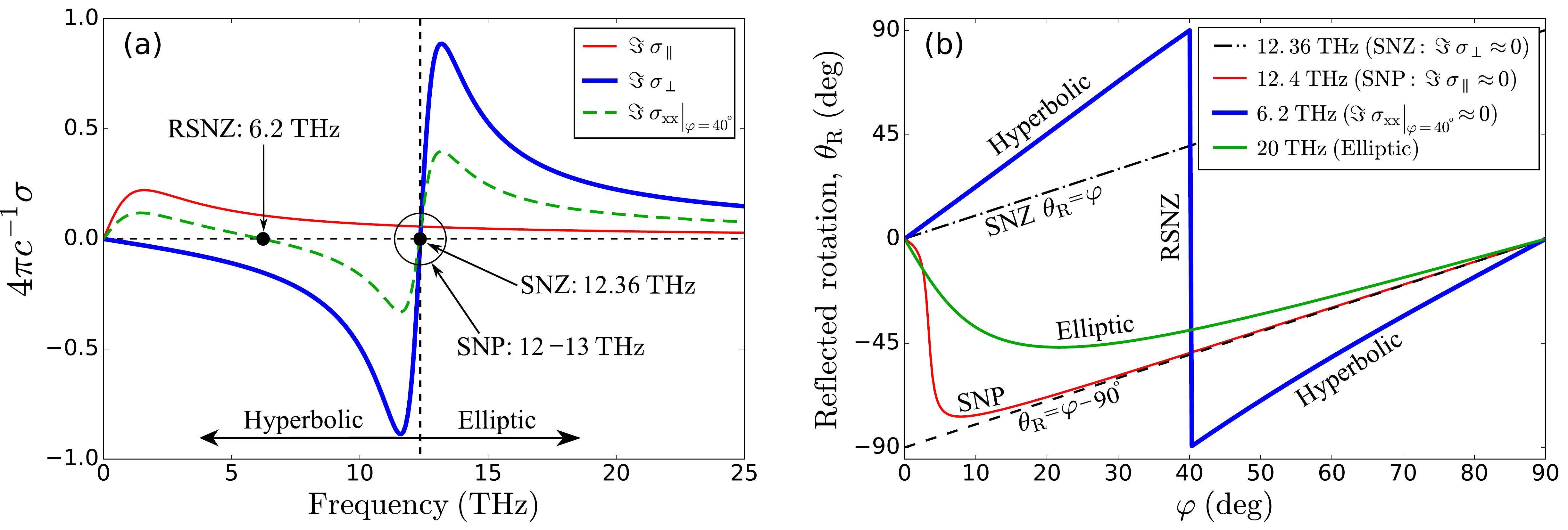} 
	\caption{\label{FigS1} (a) Imaginary parts of the effective conductivities along the strips ($\sigma_{\parallel}$), across the strips ($\sigma_{\perp}$), and in the direction along $\varphi=40^\circ$ to the strips ($\sigma_{xx}$). The black circle denotes the $\sigma$-near-pole (SNP) regime, the black points denote the $\sigma$-near-zero (SNZ) and rotated-$\sigma$-near-zero (RSNZ) regimes. (b) The reflected rotation angle $\theta_{\rm R}$ vs. the extrinsic chirality angle $\varphi$. The SNZ regime results in the linear dependence $\theta_{\rm R}=\varphi$ confirming Eq.~(\ref{th_SNZ}). The SNP regime gives the curve tending to the line $\theta_{\rm R}=\varphi-\pi/2$ confirming Eq.~(\ref{th_SNP}). The RSNZ regime results in the resonance $\theta_{\rm R}\longrightarrow\pm\pi/2$ which always occurs in the hyperbolic regime where the rotation is much stronger than in the elliptic one.    
	}	
\end{figure*}

Here we analyze the influence of the $\sigma$-near-pole (SNP) and $\sigma$-near-zero (SNZ) resonances on the reflected rotation angle $\theta_{\rm R}$ of the normally incident $p$-polarized light twisting by an achiral nonmagnetic uniaxial metasurface in the extrinsic chirality arrangement. In the main part of the article we obtained that this angle can be approximated by
\begin{align} 
	\theta_{\rm R}\simeq\arctan\left(\im\sigma_{xy}/\im\sigma_{xx}\right)=\arctan\left(\frac{\left(\im \sigma_{\parallel}-\im \sigma_{\perp}\right)\sin{\varphi}\cos{\varphi}}{\im \sigma_{\parallel}\cos^2{\varphi} +\im \sigma_{\perp}\sin^2{\varphi}}\right), \label{th_Rapp}
\end{align}		 
where $\sigma_{\parallel}$ and $\sigma_{\perp}$ are the effective conductivities along and across the main axis of a metasurface, respectively, and $\varphi$ is the angle between the plane of incidence and the main axis (extrinsic chirality angle). In the main text we have clarified that inside the SNP regime there are two SNZ regimes: exactly at the resonance frequency $\im\sigma_{\perp}=0$ (SNZ for $\sigma_{\perp}$), but near it $\im\sigma_{\parallel}\ll\im\sigma_{\perp}$ (SNZ for $\sigma_{\parallel}$). In the SNZ regime when $\im \sigma_{\perp}\approx0$ Eq.~(\ref{th_Rapp}) gives: 
\vskip -1em 
\begin{align}
	\theta_{\rm R}\approx\arctan\left(\frac{\im \sigma_{\parallel}\sin{\varphi}\cos{\varphi}}{\im \sigma_{\parallel}\cos^2{\varphi}}\right)=\arctan\left(\tan\varphi\right)=\varphi.
	\label{th_SNZ}
\end{align}
However, in the SNZ regime when $\im \sigma_{\parallel}\approx0$ it yields: 
\begin{align}
	\theta_{\rm R}\approx\arctan\left(\frac{-\im\sigma_{\perp}\sin{\varphi}\cos{\varphi}}{\im \sigma_{\perp}\sin^2{\varphi}}\right)=\arctan\left(-\cot\varphi\right)=\varphi-\pi/2.
	\label{th_SNP}
\end{align}
The rotated-$\sigma$-near-zero (RSNZ) regime, caused by the extrinsic chirality, when  $\im\sigma_{xx}\approx0$ according to Eq.~(\ref{th_Rapp}) results in the reflected rotation resonance: $\theta_{\rm R}\longrightarrow\pm\pi/2$. Using an example of the graphene strips hyperbolic metasurface we have illustrated the above approximations in Fig.~\ref{FigS1}. 
\vskip -1em 

\section{Full angle distribution} 
\begin{figure*}[h]
	\centering
	\includegraphics[scale=0.4]{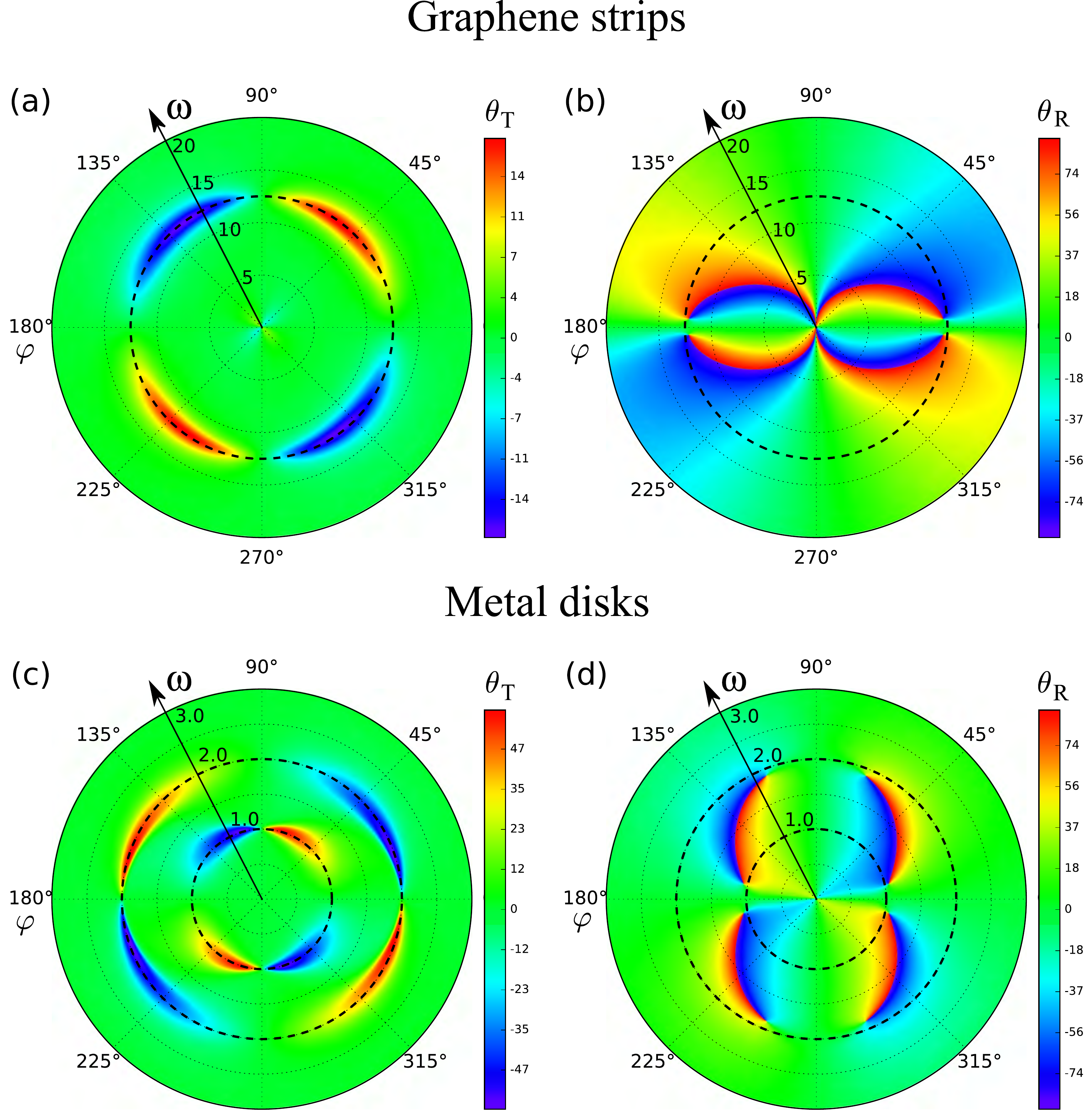} 
	\caption{\label{FigS2} Full angle distribution of $\theta_{\rm T}$, $\theta_{\rm R}$ (all in degrees) for the graphene strips (a), (b) and metal disks hyperbolic metasurfaces (c), (d). The extrinsic chirality angle $\varphi$ is the angle between the plane of incidence and the main axis of the metasurfaces, $\theta_{\rm T}$ and $\theta_R$ are the angles of the polarization plane rotation of the transmitted and reflected light, correspondingly. The radial axis corresponds in (a), (b) to frequency in THz and in (c), (d) to normalized frequency.
	}
\end{figure*}

Unlike magneto-optical Faraday and Kerr effects, where the rotation can be accumulated in the multipass experiment, in the considered achiral nonmagnetic metasurfaces after their repeated interactions with light the optical activity can be periodically compensated: the angles of the polarization plane rotation of the transmitted ($\theta_{\rm T}$) and reflected ($\theta_{\rm R}$) light will be added to $\varphi$ (the angle  between the plane of incidence and the main axis of the metasurfaces) making it sometimes lie in the first or third quadrants and sometimes in the second or fourth quadrants where the sign of rotation is opposite (Fig.~\ref{FigS2}). 
\vspace {1em}
\end{document}